\documentclass[aps,prl,twocolumn,amsmath,amssymb,superscriptaddress]{revtex4-1}
\usepackage{graphicx}
\usepackage{dcolumn}
\usepackage{bm}
\usepackage{hyperref}

\begin{document}


\title{Graphene field effect transistors with ferroelectric gating}



\author{Yi Zheng}
   \thanks{These authors contribute equally.}
   \affiliation{Department of Physics, 2 Science Drive 3, National University of Singapore, Singapore 117542}
   \affiliation{NanoCore, 4 Engineering Drive 3, National University of Singapore, Singapore 117576}
\author{Guang-Xin Ni}
   \thanks{These authors contribute equally.}
   \affiliation{Department of Physics, 2 Science Drive 3, National University of Singapore, Singapore 117542}
   \affiliation{Institute of Material Research and Engineering (IMRE),
3 Research Link, Singapore 117602}
\author{Chee-Tat Toh}
   \affiliation{Department of Physics, 2 Science Drive 3, National University of Singapore, Singapore 117542}
\author{Chin-Yaw Tan}
   \affiliation{Institute of Material Research and Engineering (IMRE),
3 Research Link, Singapore 117602}
\author{Kui Yao}
   \affiliation{Institute of Material Research and Engineering (IMRE),
3 Research Link, Singapore 117602}
\author{Barbaros \"{O}zyilmaz}
 \email{phyob@nus.edu.sg}
\affiliation{Department of Physics, 2 Science Drive 3, National University of Singapore, Singapore 117542}
   \affiliation{NanoCore, 4 Engineering Drive 3, National University of Singapore, Singapore 117576}



\begin{abstract}
Recent experiments on ferroelectric gating have introduced a novel functionality, i.e. non-volatility, in graphene field effect transistors. A comprehensive understanding in the non-linear, hysteretic ferroelectric gating and an effective way to control it are still absent. In this letter, we quantitatively characterize the hysteretic ferroelectric gating using the reference of an independent background doping ($n_\mathrm{BG}$) provided by normal dielectric gating. More importantly, we prove that $n_\mathrm{BG}$ can be used to control the ferroelectric gating by unidirectionally shifting the hysteretic ferroelectric doping in graphene. Utilizing this electrostatic effect, we demonstrate symmetrical bit writing in graphene-ferroelectric FETs with resistance change over 500\% and reproducible no-volatile switching over $10^{5}$ cycles.
\end{abstract}


\maketitle
The electric field effect, which continuously tunes the Fermi level (E$_\mathrm{F}$) in the conical energy band structure of graphene, plays a critical role in studying the extraordinary electronic properties of graphene \cite{Novoselov07NatMater,*Neto08RevModernPhys}. Using conventional dielectrics such as SiO$_2$ and more recently HfO$_2$, PMMA and Al$_2$O$_3$ with linear dielectric response to electric field, many fascinating physics have been discovered. Among these celebrated phenomena are the anomalous quantum Hall effect \cite{Novoselov05Nature,*Kim05Nature}, Klein tunneling \cite{Novoselov06NatPhys,*Kim09NaturePhys}, and gate-tunable bandgap in bilayer graphene \cite{Novoselov07PRL,*ZhangYB09Nature}. Despite such tremendous progress, there is a keen interest in the science community to utilize new dielectrics and substrates for exploring new graphene physics and functionalities \cite{Gonzalez00PRLVHS,Hatsugai06PRB,Gonzalez08PRBSuperconductivity,Valenzuela09NJPhys,Neto09PRLStrain}. Among promising candidates, ferroelectrics are unique both in ultra high dielectric constants ($\kappa$) up to a few thousands and non-linear, hysteretic dielectric response to electric field. The ultra high $\kappa$ makes ferroelectrics promising substrates for studying charge scattering mechanism in graphene \cite{Ando06JPSJ,Fuhrer08NanoTechRIP,Zhu09PZTPRL,Geim09HighKPRL}, which could be a crucial step in realizing ultra-high mobility \cite{BolotinKim08PRL,*Andrei08NatureNanotech} in non-suspended graphene. Equally important, the ultra-high $\kappa$ may allow ultra high doping in graphene with charge densities ($>10^{14}\,\mathrm{cm^{-2}}$) exceeding electrolyte doping \cite{Das08NatureNanotech} \textit{and} with gate tunability at cryogenic temperatures. Based on the hysteretic ferroelectric gating, a novel functionality of non-volatile graphene-ferroelectric field effect transistors (GFeFETs) has been demonstrated \cite{Zheng09APLGrapheneMemory}.

However, the fundamental understanding of ferroelectric gating is still elusive. In contrast to the linear doping vs normal dielectric gating relation, $n=\alpha V\mathrm{_{g}}$\cite{Novoselov07NatMater}, ferroelectric gating introduces a pronounced hysteresis in the charge doping. In particular, polymer ferroelectric gating introduces strong electron-hole puddles in graphene even far away from the Dirac point. Therefore, Hall measurements alone may be misleading in determining the induced charge doping. Thus, a quantitative modeling will not only improve the understanding of ferroelectric gating but also help in optimizing the performance of GFeFETs. Ferroelectric gating is also characterized by two symmetrical remnant polarizations, i.e. P$_{\uparrow}=\mathrm{P}_\mathrm{r}$ and P$_{\downarrow}=-\mathrm{P}_\mathrm{r}$ for upwards and downwards dipole configurations, respectively. Consequently, P$_{\uparrow}$ and P$_{\downarrow}$ induce two \textit{identical} zero-field resistance states in graphene. Although two distinct resistance states can be created by polarizing ($\mathrm{R_{0}(P_{r})}$) and depolarizing ($\mathrm{R_{1}(P\approx0)}$) the ferroelectric thin film alternately \cite{Zheng09APLGrapheneMemory}, the depolarization state is not in thermodynamic equilibrium and less stable than the polarization state. To solve this problem, we need an effective way in controlling the hysteretic ferroelectric doping. Last but not least, GFeFETs in our earlier work \cite{Zheng09APLGrapheneMemory} is characterized by low charge carrier mobility of few hundred cm$^{2}$V$^{-1}$s$^{-1}$. Such low mobility prevents the determination of the intrinsic physical properties and limitations of GFeFETs.

In this letter, we present a quantitative understanding of high quality graphene devices under ferroelectric gating. For this purpose, we introduce an independent reference doping ($n_\mathrm{BG}$) by the SiO$_2$ back gating. We show that the evolution of the device resistance hysteresis from symmetrical double peak to asymmetrical single peak structures can be consistently simulated by the electric displacement continuity equation using the reference of the SiO$_2$ gating. We also show that by controlling the polarity and magnitude of $n_\mathrm{BG}$, the hysteretic ferroelectric doping in graphene can be shifted unidirectionally. In analogy to exchange biased spin valves \cite{Nogues05PhysRep}, this effect provides a reference point for maximizing the resistance change at zero electric field and enables symmetrical bit writing in GFeFETs. We demonstrate highly reproducible non-volatile switching over $10^{5}$ cycles and $\triangle \mathrm{R/R}$ exceeding 500\% in GFeFETs.

\begin{figure}
\begin{center}
\includegraphics[width=3.4in]{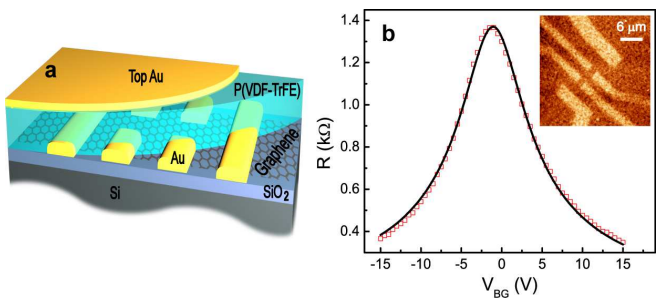}
\end{center}
\caption{(a) Sample geometry of GFeFETs. (b) R vs $V_\mathrm{BG}$ of one sample after PVDF coating. Red open square and black solid line are the experimental and fitting results respectively. Inset: Atomic force microscopy of the sample after PVDF. Color scale: 0 to 164 nm.} \label{Fig01}
\end{figure}

The GFeFET sample geometry is shown in Fig. \ref{Fig01}a. Detailed sample fabrication procedures have been discussed in Ref. \cite{Zheng09APLGrapheneMemory}. For the samples used in this study, the ferroelectric thin film of poly(vinylidene fluoride-trifluoroethylene 72:28) (PVDF) is $\sim0.5\,\mu$m thick. The GFeFETs were electrically characterized at room temperature in vacuum using four-contact lock-in technique.

Before polarizing the ferroelectric, we first measured the Hall mobility and the resistance vs SiO$_2$ gate voltage characteristics (R vs $V_\mathrm{BG}$) to determine the sample quality. Most samples retain their high mobility after PVDF spin-coating and annealing, as shown in Fig. \ref{Fig01}b for a typical sample with Hall mobility of 4,600 cm$^{2}$V$^{-1}$s$^{-1}$ \cite{Note01}. Quantitatively, the ambipolar R vs $V_\mathrm{BG}$ characteristics can be fitted very well by the model \cite{Kim08APLAl2O3},
\begin{equation}\label{Equ01}
   \mathrm{R}=\frac{L}{We\mu_{Hall}\sqrt{n_\mathrm{res}^2+n^2}},
\end{equation}
using the Hall mobility $\mu_{Hall}$. For the sample shown in Fig. \ref{Fig01}b, the fitting yields a residual carrier concentration $n_\mathrm{res}=2.77\times10^{11}\,\mathrm{cm^{-2}}$.

Compared to the SiO$_2$ gating, one fundamental difference introduced by ferroelectric gating is pronounced hysteresis in the resistance vs ferroelectric gate voltage characteristics (R vs $V_\mathrm{TG}$). Though such hysteretic R vs $V_\mathrm{TG}$ can be qualitatively explained by the electric displacement continuity equation at the ferroelectric/graphene interface \cite{Zheng09APLGrapheneMemory}, a quantitative understanding of ferroelectric gating is still missing. Here, we introduce an independent $n_\mathrm{BG}$ using the SiO$_2$/Si back gate. This provides a well defined, constant reference for determining the doping induced by PVDF gating. To study the effect of $n_\mathrm{BG}$ on the ferroelectric gating of GFeFETs, it is also important to limit the polarization magnitude in PVDF, since the effect of ferroelectric gating is nearly 10 times stronger than the SiO$_2$ gating \cite{Zheng09APLGrapheneMemory}. Thus, we first introduced very small $|\mathrm{P}_\mathrm{r}|$ in PVDF by limiting the maximum top gate voltage ($V_\mathrm{TGmax}$) to $\pm5$ V. Such low $V_\mathrm{TG}$ only slightly polarizes PVDF, allowing $n_\mathrm{BG}$ to match or even exceed the $|\mathrm{P}_\mathrm{r}|$ induced doping in graphene.

\begin{figure*}
\begin{center}
\includegraphics[width=4.7in]{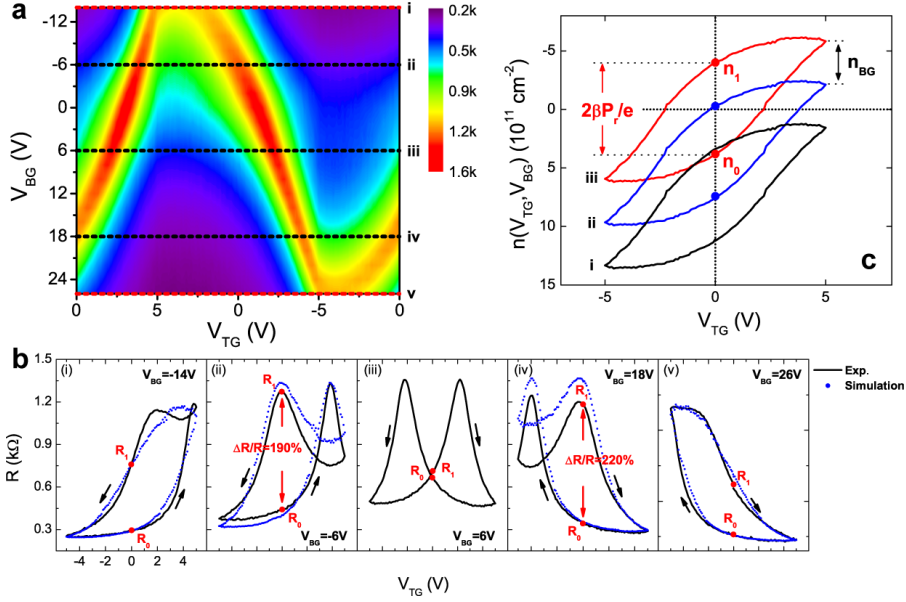}
\end{center}
\caption{(a) R vs $V_\mathrm{TG}$ and $V_\mathrm{BG}$ of the GFeFET with very small $\mathrm{|P_{r}}|$. (b) Extracted single traces of R vs $V_\mathrm{TG}$ with different $V_\mathrm{BG}$. The blue dotted lines are simulated results. (c) $n_{BG}$ tunable doping hysteresis in GFeFETs. Only three doping hysteresis loops, corresponding to experimental curves i, ii and iii in Fig. \ref{Fig02}b, are shown for clarity. See main texts for discussions.}  \label{Fig02}
\end{figure*}

In Fig. \ref{Fig02}a, we show the resistance of the GFeFET as a function of both $V_\mathrm{TG}$ and $V_\mathrm{BG}$. With $V_\mathrm{BG}\approx6$ V, the R vs $V_\mathrm{TG}$ curve shows two symmetrical resistance peaks and nearly negligible $\triangle \mathrm{R/R}$ (Fig. \ref{Fig02}b(iii)). By gradually tuning $n_\mathrm{BG}$ with $V_\mathrm{BG}$, the two resistance peaks become more asymmetrical and shift leftward (rightward) for $n_\mathrm{BG}<0$ ($n_\mathrm{BG}>0$). The shift in peak positions leads to an increase in $\triangle \mathrm{R/R}$, which has a maximum at $V_\mathrm{BG}\approx-6$ V (Fig. \ref{Fig02}b(ii)) and $V_\mathrm{BG}\approx18$ V (Fig. \ref{Fig02}b(iv)), respectively. Crossing these two points, $\triangle \mathrm{R/R}$ decreases as $|n\mathrm{_{BG}}|$ keeps on increasing. At large enough $n_\mathrm{BG}$, the double peak structure eventually disappears in the R vs $V_\mathrm{TG}$ hysteresis (Fig. \ref{Fig02}b(i) and Fig. \ref{Fig02}b(v)).

The evolution of the resistance peaks and the change in $\triangle \mathrm{R/R}$ can both be explained by two independent but competing doping processes in graphene by polarized ferroelectric dipoles and $V_\mathrm{BG}$, respectively. For such a dual-gated system, the interfacial electric displacement continuity equation is expressed by
\begin{equation}\label{Equ02}
   -\beta\mathrm{P}(V_\mathrm{TG})+n^{\ast}=n(V_\mathrm{TG}, V_\mathrm{BG})e,
\end{equation}
where $\beta\mathrm{P}(V_\mathrm{TG})$ represents the hysteretic dipole doping by the ferroelectric gating \cite{Note02}, and $n^{\ast}=n_\mathrm{env}+n_\mathrm{BG}$ is the reference doping induced by the dielectric environment and $V_\mathrm{BG}$ respectively. For $n^{\ast}\approx0$, the doping in graphene is dominated by the ferroelectric gating by $n(V_\mathrm{TG}, n^{\ast}\approx0)=-\beta\mathrm{P}(V_\mathrm{TG})/e$. Using Eq. \ref{Equ01}, it is now straightforward to see that $n(V_\mathrm{TG}, n^{\ast}\approx0)$ will produce a R vs $V_\mathrm{TG}$ hysteresis with two symmetrical resistance peaks, centering on the two coercive-field points where $\mathrm{P}(V_\mathrm{TG})$ crossing zero. Experimentally, this is the R vs $V_\mathrm{TG}$ curve in Fig. \ref{Fig02}b(iii) with $V_\mathrm{BG}=6$ V, in which two resistance peaks are centered at $V_\mathrm{TG}=\pm2.2\,\mathrm{V}$ respectively. By converting each R in Fig. \ref{Fig02}b(iii) into doping using Eq. \ref{Equ01}, we directly determined the doping curve $n(V_\mathrm{TG},n^{\ast}\approx0)$. The result is shown in Fig. \ref{Fig02}c (red curve). As expected, this doping curve is hysteretic and characterized by two zero-field doping levels with equal magnitude, i.e. $|n_{1}|=|n_{0}|=\beta\mathrm{P_{r}}/e$.

After acquiring $n(V_\mathrm{TG},n^{\ast}\approx0)$, we can deduce individual $\mathrm{R}(V_\mathrm{TG}, n^{\ast})$ curves for non-zero $n^{\ast}$ by substituting $n(V_\mathrm{TG},V_\mathrm{BG})=-\beta\mathrm{P}(V_\mathrm{TG})/\mathrm{e}+n_\mathrm{env}+\alpha V_\mathrm{BG}$ into Eq. \ref{Equ01}. Here $\alpha=7.2\times10^{10}\,\mathrm{cm^{-2}V^{-1}}$ is the doping coefficient of 300 nm SiO$_2$, and $n_\mathrm{env}$ is a fitting parameter \cite{Note03}. By tuning $n_\mathrm{env}$ and matching the resistance peaks of the simulation to the experimental, we simulated each experimental R($V_\mathrm{TG}$, $V_\mathrm{BG}$) curve in Fig. 2b. As shown by blue dotted lines, the simulation reproduces the evolution of the experimental results very well. Two resulting doping hysteresis for the resistance curves in Fig. \ref{Fig02}b(i) and Fig. \ref{Fig02}b(ii) are further compared with $n(V_\mathrm{TG},n^{\ast}\approx0)$ in Fig. \ref{Fig02}c. From the comparison, we can see that $\triangle \mathrm{R/R}$ approaches the maxima as one zero-field doping level sits near the Dirac point when $|n^{\ast}|\approx \beta\mathrm{P}_\mathrm{r}/e$ (blue hysteresis loop). Further increase in $n_\mathrm{BG}$ moves \textit{both} $n_{1}$ and $n_{0}$ away from the Dirac point, and $\triangle \mathrm{R/R}$ decreases (black hysteresis loop).

\begin{figure}
\begin{center}
\includegraphics[width=3.5in]{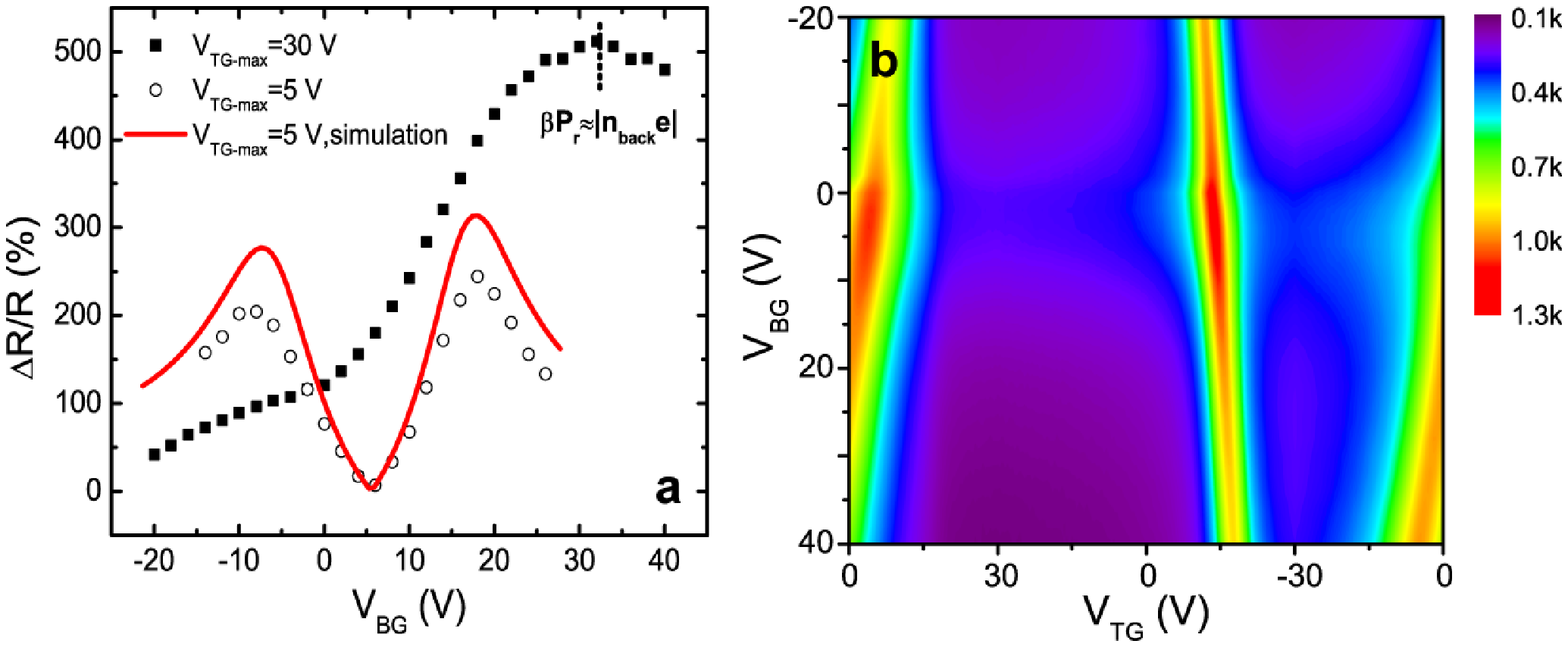}%
\end{center}
\caption{\label{Fig03}(a) $\triangle \mathrm{R/R}$ as a function of $V_\mathrm{BG}$ with different $V_\mathrm{TGmax}$. Two maxima are observable with V$_\mathrm{TGmax}=5$ V (black open circles). The red solid line shows the simulation with $\beta\mathrm{P_{r}}=4.2\times10^{11}\,\mathrm{cm^{-2}}$. For V$_\mathrm{TGmax}=30$ V, the maximum $\triangle \mathrm{R/R}$ is increased to 500\% (V$_\mathrm{BG}=32$ V). (b) R (V$_\mathrm{TG}$, V$_\mathrm{BG}$) of the GFeFET with higher $\beta\mathrm{P}_\mathrm{r}$ ($\sim2\times10^{12}\,\mathrm{cm^{-2}}$). Double peak structures dominate over the whole V$_\mathrm{BG}$ range. }
\end{figure}

Thus, we have shown that using a background doping introduced by normal dielectric gating as a reference, the hysteretic behavior of R vs ferroelectric gating in GFeFETs can be quantitatively determined by solving the electric displacement continuity equation. For memory applications, $\triangle \mathrm{R/R}$ is of great importance. Following the above discussions, the two zero-field resistance states are $\mathrm{R_{1}}=\frac{L}{We\mu\sqrt{n_\mathrm{res}^2+(\beta\mathrm{P_{r}/e}-n^{\ast})^2}}$ and $\mathrm{R_{0}}=\frac{L}{We\mu\sqrt{n_\mathrm{res}^2+(\beta\mathrm{P_{r}/e}+n^{\ast})^2}}$ respectively. Thus, the best strategy to utilize the field-dependent resistance is to fully polarize the ferroelectric and introduce a matching $n_\mathrm{BG}$, as demonstrated in Fig. \ref{Fig03}a. With $V_\mathrm{TGmax}=5$ V, two maxima of $\sim$250\% are present in $\triangle \mathrm{R/R}$ vs $V_\mathrm{BG}$, which can be also simulated very well by Eq. \ref{Equ01} and \ref{Equ02} with $\beta\mathrm{P_{r}}/e=4.2\times10^{11}\,\mathrm{cm^{-2}}$. By increasing $V_\mathrm{TGmax}$ to 30 V, the maximum $\triangle \mathrm{R/R}$ is increased to 500\%. The fast increase in $\mathrm{P_{r}}$ not only increases the maximum $\triangle \mathrm{R/R}$, but also increases the separation between the two $\triangle \mathrm{R/R}$ maxima, resulting in one maximum being outside of $V_\mathrm{BG}$ measurement range. For this $V_\mathrm{TGmax}$, R vs $V_\mathrm{TG}$ shows a dominant double peak structure over the full $V_\mathrm{BG}$ range (Fig. \ref{Fig03}b). However, we can still see the tendency of a transition from double peak structure to single peak structure as $V_\mathrm{BG}$ exceeding 40 V.

\begin{figure}
\begin{center}
\includegraphics[width=3.4in]{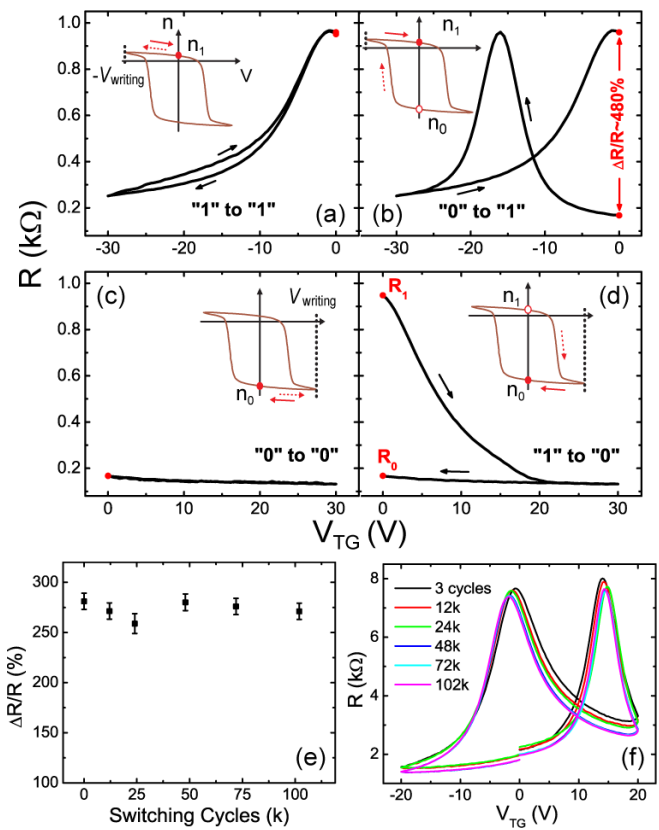}%
\end{center}
\caption{\label{Fig04} Symmetrical bit writing in GFeFETs with $n_\mathrm{BG}\approx -\beta \mathrm{P_{r}}$/e. (a) and (b) Writing ``1'' using $-V_\mathrm{writing}$. (c) and (d) Writing ``0'' using $V_\mathrm{writing}$. Dashed and solid arrows indicate the forward and backward voltage sweep directions respectively. The writing is independent on the initial states. (e) and (f) Fatigue test of one GFeFET with symmetrical bit writing, showing non-volatile switching cycles exceeding 100k.}
\end{figure}

Such $n_\mathrm{BG}$ shifted hysteretic doping in graphene is a ferroelectric analogy to the ferromagnetic exchange bias \cite{Nogues05PhysRep}. Utilizing this electrostatic effect, the bit writing in GFeFETs can be much simplified by switching the ferroelectric polarization between P$_\mathrm{r}$ and -P$_\mathrm{r}$, using symmetrical voltage sweeps. With $n_\mathrm{BG}\approx -\beta \mathrm{P_{r}}$/e, to write the high resistance ``1'', a negative writing voltage (-$V_\mathrm{writing}$) is applied to the ferroelectric, setting the dipole polarization to -P$_\mathrm{r}$ independent of the initial states in the unit cell (Fig. \ref{Fig04}a and \ref{Fig04}b). In contrast, a positive $V_\mathrm{writing}$ with the same magnitude sets the GFeFET into low resistance ``0'' (Fig. \ref{Fig04}c and \ref{Fig04}d). Compared to the asymmetrical bit writing by polarizing and depolarizing the ferroelectric alternately \cite{Zheng09APLGrapheneMemory}, such symmetrical writing in GFeFETs not only provides simplicity but also takes full advantage of the fast switching speed of ferroelectric. For lead zirconate titanate (PZT) based materials, this can be as fast as 280 ps \cite{Scott05RevModernPhys}. Another potential application of this electrostatic effect could be \textit{multi-bit-per-cell} data storage in GFeFETs utilizing the $n_\mathrm{BG}$ tunable $\triangle \mathrm{R/R}$.

We have also tested the reproducibility of our GFeFETs working with $\beta\mathrm{P_{r}}\approx |n^{\ast}|e$. During the fatigue test, a triangular wave of 1k Hz was applied to the PVDF thin film. Every 12 (24) seconds, the triangular wave was interrupted and one R vs $V_\mathrm{TG}$ curve was recorded. The corresponding $\triangle \mathrm{R/R}$ as a function of switching cycles and the raw data of individual R vs $V_\mathrm{TG}$ curve are summarized in Fig. \ref{Fig04}e and \ref{Fig04}f respectively. The fatigue test clearly demonstrates reproducible non-volatile switching exceeding 100k cycles in the GFeFET. Ultimately, the life span of PVDF-based GFeFETs is $10^{7}$ \cite{Furukawa06IEEE}. Thus, PVDF-GFeFETs could provide a cost-effective solution for flexible non-volatile data storage with sub-$\mu$s switching speed. On the other hand, inorganic ferroelectric (such as PZT) should be used if fast writing speed ($<$ ns) and ultra-high endurance ($10^{10}$) are required.

In conclusion, we have demonstrated a quantitative way in determining and controlling the hysteretic ferroelectric gating in GFeFETs. Using an independent linear dielectric gating ($n_\mathrm{BG}$) as a reference \cite{Note04}, ferroelectric gating can be quantitatively determined by the electric displacement continuity equation. The reference gating can also be used to control ferroelectric gating by introducing a unidirectional shift in the hysteretic ferroelectric doping in GFeFETs. One specific application of this electrostatic ``bias'' effect is symmetrical bit writing in GFeFETs directly utilizing $\mathrm{P_{r}}$ and $-\mathrm{P_{r}}$ with speed and simplicity. The ferroelectric gating phenomena and related modeling and controlling methods presented in this study will be important in understanding future charge transport studies on ferroelectric gated graphene electronic devices.

\begin{acknowledgements}
We particularly acknowledge Douwe J. Monsma for many insights and useful discussions. This work is supported by the Singapore National Research Foundation (NRF-RF2008-07), NRF-CRP grant Graphene and Related Materials and Devices (R-143-000-360-281), NUS SMF Award, ONR Award, and by NUS NanoCore.
\end{acknowledgements}


\end{document}